\documentclass[
superscriptaddress,
nofootinbib, twocolumn,
amsmath,amssymb,aps,
floatfix, eprint ]{revtex4-2}

\usepackage[T1]{fontenc} 

\usepackage{graphicx}
\usepackage{dcolumn}
\usepackage{bm}
\usepackage{epstopdf}
\usepackage{mathrsfs}
\usepackage{amsmath,amssymb,amsfonts,latexsym}
\usepackage{mathtools}
\usepackage{amsmath,bbold}
\usepackage{color}
\usepackage{slashed}
\usepackage{nicefrac}
\usepackage[colorlinks=true,linktocpage=true,linkcolor=blue,citecolor=blue]{hyperref}
\usepackage[utf8]{inputenc}
\usepackage[T1]{fontenc}
\usepackage{accents}
\usepackage[normalem]{ulem}
\usepackage{cancel}
\usepackage{tikz}
\usepackage{cleveref}

\long\def\comment#1{ }

\def\0{{\boldsymbol 0}}

\def\k{{\boldsymbol k}}

\def\x{{\boldsymbol x}}

\def\p{{\boldsymbol p}}

\def\v{{\boldsymbol v}}

\def\and{\qquad\text{and}\qquad}

\def\min{\text{min}}

\newcommand{\beq}{\begin{eqnarray}}
\newcommand{\eeq}{\end{eqnarray}}
\newcommand{\be}{\begin{eqnarray*}}
\newcommand{\ee}{\end{eqnarray*}}
\newcommand{\bal}{\begin{align}}
\newcommand{\eal}{\end{align}}
\newcommand{\rmd}{{\rm d}}

\begin{document}
\title{Deriving a parton shower for jet thermalization in QCD plasmas}

\author{Ismail Soudi}
\email{isma@physik.uni-bielefeld.de}
\affiliation{University of Jyväskylä, Department of Physics, P.O. Box 35, FI-40014 University of Jyväskylä, Finland.}
\affiliation{Helsinki Institute of Physics, P.O. Box 64, FI-00014 University of Helsinki, Finland.}
\affiliation{Fakult\"at f\"ur Physik, Universit\"at Bielefeld, D-33615 Bielefeld, Germany}

\author{Adam Takacs}
\email{takacs@thphys.uni-heidelberg.de}
\affiliation{Institute for Theoretical Physics, Heidelberg University, Philosophenweg 16, 69120 Heidelberg, Germany}
\begin{abstract}
    Jet quenching---the modification of high-energy jets in the quark–gluon plasma---has been extensively studied through weakly coupled scattering amplitudes embedded in parton-shower frameworks. These models, often combined with bulk hydrodynamic evolution, successfully describe a wide range of observables, though they typically rely on assumptions of rapid thermalization and simplified treatments of medium response. Parallel to these developments, jet thermalization has been investigated within the finite-temperature QCD effective kinetic theory, which provides our best microscopic understanding of equilibration in heavy-ion collisions. Early studies of linearized perturbations have highlighted both the promise and the limitations of current approaches, as existing MC implementations face challenges---particularly in the treatment of recoils and particle merging. Building on this foundation, we introduce a new parton-shower algorithm that exactly reproduces the dynamics of the linearized EKT, enabling a first-principles description of jet thermalization with proper inclusion of recoils, holes, quantum statistics, and merging processes.
\end{abstract}
\maketitle

\section{Introduction}
Jet quenching is one of the defining signatures of the quark–gluon plasma (QGP) formed in high-energy heavy-ion collisions~\cite{Busza:2018rrf,Harris:2023tti}. In this hot, dense, and color-deconfined medium, energetic partons and jets undergo significant modification relative to proton–proton collisions~\cite{Armesto:2015ioy,Connors:2017ptx,Cunqueiro:2021wls}, providing key insights into the microscopic properties of the QGP~\cite{JET:2013cls,JETSCAPE:2021ehl,JETSCAPE:2024cqe}.

In the weak-coupling regime, jet quenching arises from medium-induced modifications of scattering amplitudes~\cite{Casalderrey-Solana:2007knd,Majumder:2010qh,Mehtar-Tani:2013pia,Qin:2015srf,Mehtar-Tani:2025rty}, which are implemented in parton-shower Monte Carlo (MC) frameworks~\cite{Majumder:2013re,Zapp:2013vla,Caucal:2018dla,Blanco:2020uzy}. These approaches successfully reproduce a broad range of jet cross sections and substructure observables~\cite{CMS:2021vui,ATLAS:2022vii,ATLAS:2023iad,ALICE:2023waz,ALICE:2024fip,CMS-PAS-HIN-24-016}. The state-of-the-art jet quenching models separate early vacuum-like jet evolution from medium-modifications. In this work, we focus on the latter and on the approach to equilibrium.

Recent developments aim to couple jet evolution to realistic, hydrodynamically evolving QGP backgrounds~\cite{Schenke:2010rr,Nijs:2020ors,Wu:2021fjf}, requiring a consistent treatment of energy–momentum deposition and medium response~\cite{Cao:2020wlm,He:2022evt,Barreto:2022ulg,Cao:2024pxc,vanderSchee:2025hoe}. Typically, these models impose an infrared cutoff of $\approx\!5$ GeV, below which partons are assumed to equilibrate instantaneously and to source hydrodynamic wakes~\cite{Cao:2020wlm,Cao:2024pxc}. It is not clear, however, whether partons are indeed equilibrated in this regime. Analogous approaches in strongly coupled frameworks support rapid equilibration~\cite{Gubser:2007ga,Chesler:2007sv,Casalderrey-Solana:2020rsj,Pablos:2022piv}. Although such models have achieved notable phenomenological success~\cite{CMS:2021otx,CMS:2025dua}, they remain phenomenological in nature and are not derived from first principles.

In contrast, the finite-temperature QCD effective kinetic theory (EKT)~\cite{Arnold:2002zm} provides a first-principles description of thermalization in the weak-coupling limit, consistent with the ``bottom-up'' scenario~\cite{Baier:2000sb}---our best microscopic understanding of thermalization in heavy-ion collisions~\cite{Schlichting:2019abc,Berges:2020fwq}. Extending this framework to jet thermalization is therefore a natural next step. Initial progress has been made through studies of the equilibration of small perturbations (``minijets'')~\cite{Schlichting:2020lef,Mehtar-Tani:2022zwf,Zhou:2024ysb}. These studies revealed significant deviations from equilibrium in the phenomenologically relevant few-GeV region by solving the linearized Boltzmann equation numerically. This formulation, however, is not easily interfaced with event-by-event phenomenological workflows that require particle-level outputs for jet reconstruction, hadronization, and comparison with experimental data.

The non-linear Boltzmann equation has been solved using stochastic approaches in MC cascades for the heavy-ion bulk~\cite{Zhang:1997ej,Bass:1998ca,Lin:2004en,SMASH:2016zqf,Kurkela:2022qhn}. These implementations focus on the bulk medium and face difficulties with high-energy excitations due to the large separation of scales between jet and medium particles. Alternatively, attempts to adapt the linearized Boltzmann equation into MC simulations~\cite{Zapp:2008gi,Auvinen:2009qm,Schenke:2009gb,He:2015pra,He:2018xjv} face their own limitations. These include approximate treatments of jet-medium interactions through ``recoil'' and ``hole'' particles, the neglect of secondary collisions, the omission of $2\!\to\!1$ merging processes, and the absence of quantum statistical effects. Such simplifications prevent proper equilibration and thus hinder a consistent description of jet thermalization.

In this work, we present a new MC algorithm that exactly reproduces the dynamics of the linearized EKT. It consistently incorporates energy loss, thermalization, Bose/Fermi statistics, the tracking of recoils and holes, and $2\!\to\!1$ particle merging. This enables a fully microscopic, first-principles description of jet energy loss and thermalization in the QGP. Furthermore, leveraging the multi-particle capabilities of MC simulations, we compute for the first time the two-particle correlation within the parton shower framework, revealing correlations beyond molecular chaos.

\section{Theoretical framework}
Following the EKT framework~\cite{Arnold:2002zm}, the evolution of a small perturbation $\delta f(t,\x,\p)$ in an equilibrium background $n(p)$ is described by the linearized Boltzmann equation~\cite{Schlichting:2020lef,Zhou:2024ysb},
\begin{equation}\label{eq:linBoltzmann}
    \left(\partial_t+\v\cdot
    \nabla_{\x}\right)\delta f_a(t,\x,\p)
    = \delta C_a^{2\leftrightarrow2} + \delta C_a^{1\leftrightarrow2}\,,
\end{equation}
where $\v\!=\!\p/p$, $\delta C_a^{2\leftrightarrow2}[\{n,\delta f\}](t,\x,\p)$ and $\delta C_a^{1\leftrightarrow2}[\{n,\delta f\}](t,\x,\p)$ describe linearized elastic and inelastic collisions, respectively, and $a$ denotes the parton flavor (quark or gluon). The equilibrium distribution for bosons and fermions is $n_a(t,\x,\p)=1/(e^{p/T}\mp1)$. Inelastic $1\leftrightarrow2$ processes are given by
\begin{equation}\label{eq:linC12}
    \begin{split}
        \delta C_a^{1\leftrightarrow2} & = \sum_{bc}\int^1_0dz\left[\tfrac{1}{z^3}\Gamma^c_{ab}\left(z,\tfrac{\bm p}{z}\right) \delta \mathcal F^c_{ab}\left(\tfrac{\p}{z};\p,\tfrac{\bar z\p}{z}\right) \right. \\
                                       & \left.- \tfrac12\Gamma^a_{bc}(z,\bm p)\delta\mathcal F^a_{bc}(\p;z\p,\bar z\p)\right] \,,
    \end{split}
\end{equation}
where $\Gamma$ is the splitting rate, and the corresponding statistical factor is
\begin{equation}\label{eq:linStatF12}
    \begin{split}
         & \delta\mathcal F^a_{bc}(\p;\k,\p')
        =                                                                                       \\
         & \,\, +\delta f_a(\p)\left[(1 \pm n_b(k))(1 \pm n_c(p')) \mp_a n_b(k)n_c(p')\right]   \\
         & \,\, - \delta f_b(\k)\left[n_c(p')(1 \pm n_a(p)) \mp_b n_a(p)(1\pm n_c(p'))\right]   \\
         & \,\, - \delta f_c(\p')\left[n_b(k)(1 \pm n_a(p)) \mp_c n_a(p)(1\pm n_b(k))\right]\,, \\
    \end{split}
\end{equation}
where $\pm_a$ sign varies for bosons and fermions and the $\x$ dependence is local and implicit. Elastic collisions are given by
\begin{equation}\label{eq:linC22}
    \begin{split}
        \delta C_a^{2\leftrightarrow2} & = \frac{1}{4|\bm p|\nu_a}\sum_{bcd}\int \rmd\Omega^{2\leftrightarrow2}|\mathcal M^{ab}_{cd}|^2 \delta \mathcal F^{ab}_{cd}\,,
    \end{split}
\end{equation}
where $\int \rmd\Omega^{2\leftrightarrow2}=\int_{k,p',k'}\delta^4(p^\mu+k^\mu-p'^\mu-k'^\mu)$, with $\int_p=\int_{\p}\frac{1}{2E_p}$, $\int_{\p}=\int\frac{d^3\p}{(2\pi)^3}$, and $|\mathcal M^{ab}_{cd}|^2$ are the HTL regulated leading-order 2-2 matrix elements, and the associated statistical factor is
\begin{equation}\label{eq:linStatF22}
    \begin{split}
         & \delta\mathcal{F}^{ab}_{cd}(\p,\k;\p',\k') =                                       \\
         & + \delta f_c(\p') \left[n_d(1 \pm n_a)(1 \pm n_b) \mp_c n_an_b(1\pm n_d)\right]    \\
         & + \delta f_d(\k') \left[n_c(1 \pm n_a)(1 \pm n_b) \mp_d n_an_b(1\pm n_c)\right]    \\
         & - \delta f_a(\p)  \left[n_b(1 \pm n_c)(1 \pm n_d)\mp_a n_cn_d(1\pm n_b)\right]     \\
         & - \delta f_b(\k)  \left[n_a(1 \pm n_c)(1 \pm n_d)\mp_b n_cn_d(1\pm n_a)\right] \,.
    \end{split}
\end{equation}
We omit the momentum dependence, which can be inferred from the indices $(a,b,c,d)$ corresponding to $(a,\p)$, $(b,\k)$, $(c,\p')$, $(d,\k')$, while the $\x$ dependence is local an implicit.

In the following, we introduce the parton shower formulation of the linearized Boltzmann equation. In this work, we neglect space-time inhomogeneities, and so $\delta f(t,\p)=\int\rmd^3\x\,\delta f(t,\x,\p)$. We first introduce the shorthand notation
\begin{equation}
    \begin{split}
        \delta C^{1\leftrightarrow2}_a & = \delta C^{1\leftrightarrow2,{\rm r}}_a[\delta f] - \delta C^{1\leftrightarrow2,{\rm v}}_a\cdot\delta f_a(t,\p)\,, \\
        \delta C^{2\leftrightarrow2}_a & = \delta C^{2\leftrightarrow2,{\rm r}}_a[\delta f] - \delta C^{2\leftrightarrow2,{\rm v}}_a\cdot\delta f_a(t,\p)\,,
    \end{split}
\end{equation}
where the right-hand side can be read off from \cref{eq:linC12,eq:linStatF12,eq:linC22,eq:linStatF22},
distinguishing real and virtual momentum exchanges. Virtual collisions do not change momentum and are therefore directly proportional to $\delta f_a(\p)$. We rewrite \cref{eq:linBoltzmann} as
\begin{equation}\label{eq:PartonShower}
    \begin{split}
        \delta f_a(t,\p) & = \Delta_a(t,\p)\delta f_{a0} + \int_{t_0}^t\rmd t'\Delta_a(t-t',\p)                                           \\
                         & \times(\delta C^{1\leftrightarrow2,{\rm r}}_a[\delta f] + \delta C^{2\leftrightarrow2,{\rm r}}_a[\delta f])\,,
    \end{split}
\end{equation}
where we used the initial condition $\delta f_{a0}=\delta f_a(t_0,\p)$, and
\begin{equation}\label{eq:Sudakov}
    \begin{split}
         & \Delta_a(t,\p) = \exp\left[-\int^t_{t_0}\rmd t' (\delta C^{1\leftrightarrow2,{\rm v}}_a + \delta C^{2\leftrightarrow2,{\rm v}}_a)\right]\,.
    \end{split}
\end{equation}
\Cref{eq:PartonShower} is the parton shower formulation of the linearized Boltzmann equation, where $\Delta_a(t,\p)$ is the no-collision probability, analogous to the Sudakov factor in vacuum jet showers~\cite{Ellis:318585}. The no-collision probability and the real-collision terms $\delta C^{1\leftrightarrow2,{\rm r}}_a$, $\delta C^{2\leftrightarrow2,{\rm r}}_a$ include splitting, merging, recoil, and hole particles, and therefore go beyond typical MC implementations of elastic~\cite{Zapp:2008gi,Auvinen:2009qm,Li:2010ts,DEramo:2018eoy} and inelastic processes~\cite{Zapp:2008af,Wang:2013cia,Caucal:2019uvr,Blanco:2020uzy}. The appearance of holes (negative particles) becomes more transparent here, as $\delta C^{,{\rm r}}$ contains negative terms. Solving \cref{eq:PartonShower} is equivalent to solving the original linearized Boltzmann equation and therefore provides a consistent description of the equilibration of perturbations in QCD plasmas. When \cref{eq:PartonShower} is solved using MC techniques, initial particles are sampled from $\delta f_{a0}$ and subsequently undergo real collision processes according to the no-emission probability and real-collision kernels.

\section{High-energy limit}
\label{sec:high-energy_limit}
In this section, we take a simplified version of \cref{eq:PartonShower} to introduce the basic concepts of parton-shower algorithms before proceeding to the general case. We also demonstrate that \cref{eq:PartonShower} reproduces the DGLAP-like energy-loss evolution, i.e., the turbulent cascade, used in most jet-quenching models.\footnote{We remind the reader that the early vacuum-like jet evolution is neglected here, and we focus on energy loss.}

For simplicity, let us consider a gluon plasma. In the high-energy, under-occupied limit, with $p\gg T$ and $n(p)\ll1$, the $1\to2$ collision integral simplifies to pure radiation:
\begin{equation}
    \begin{split}
        \delta C^{1\leftrightarrow2} & \approx \int_0^1\rmd z\left[\tfrac{1}{z^3}\Gamma(z,\tfrac{\p}{z})\delta f(t,\tfrac{\p}{z})\right. \\
                                     & -\left.\tfrac{1}{2}\Gamma(z,\p)\delta f(t,\p)\right]\,.
    \end{split}
\end{equation}
We integrate out the angle and introduce the energy density $D(t,x;p_0)=\nu_g\int_{\p}\tfrac{|\p|}{p_0}\delta(\tfrac{|\p|}{p_0}-x)\delta f(t,\p)$, arriving at the well-known turbulent cascade evolution~\cite{Blaizot:2013vha,Blaizot:2013hx,Mehtar-Tani:2018zba,Schlichting:2020lef}
\begin{equation}\label{eq:cascade}
    \begin{split}
        \partial_t D(t,x;p_0) & = \int_0^1 \rmd z\left[\Gamma\left(z,\tfrac{x\p_0}{z}\right)D\left(t,\tfrac{x}{z};\p_0\right)\right. \\
                              & -\left.\tfrac12\Gamma(z,x\p_0)D(t,x;\p_0)\right]\,,
    \end{split}
\end{equation}
where the energy fraction is $x=p/p_0$. This (or similar) evolution is used in many jet quenching MCs with different splitting rates $\Gamma$, depending on their modeling of the jet-medium interaction.

For illustration, we derive the parton-shower algorithm in this simple limit:
\begin{equation}\label{eq:cascadePS}
    \begin{split}
        D(t,x;p_0) & = \Delta(t,x\p_0)D_0 + \int_{t_0}^t\rmd t'\int_0^1\rmd z                    \\
                   & \times\Delta(t-t',x\p_0)\Gamma(z,\tfrac{x\p_0}{z})D(t',\tfrac{x}{z};p_0)\,,
    \end{split}
\end{equation}
where the no-emission probability is $\Delta(t,\p)=\exp[-\int_{t_0}^t\rmd t'\int_0^1\rmd z\tfrac12\Gamma(z,\p)]$. For a single parton initial condition, $D_0=x\delta(1-x)$, the first few splittings are
\begin{equation}\label{eq:cascadePS_exp}
    \begin{split}
         & D(t,x;p_0) = \Delta(t,x\p_0)x\delta(1-x)                                                \\
         & \quad+  \int_{t_0}^t\rmd t_1 \Delta(t-t_1,x\p_0)x\Gamma(x,\p_0)\Delta(t_1,\p_0) + \dots
    \end{split}
\end{equation}
The first line represents the case with no emissions throughout the evolution, while the second line describes exactly one emission at $t_1$, with a splitting fraction $x$, where the two Sudakov factors prevents further emissions before and after. In the parton shower algorithm, first, the splitting time is determined by sampling the Sudakov $\Delta(t_{\rm next},\p)$. In case of splitting, $t_{\rm next}<L$, two new particles are created (note the factor of 2 difference between the Sudakov's exponent and splitting rate in \cref{eq:cascadePS}) with $zp,(1-z)p$ energies, where $z$ is sampled from the rate $\tfrac12\Gamma(z,p)$. This procedure is repeated for all new particles until $t_{\rm next}>L$. A sampling of a final Sudakov ensures that particles do not split after their last creation, $\Delta(L-t_{\rm last},p')$. Energy-momentum conservation is encoded in the collinear momentum reconstruction, while the particle number is not conserved. In practice, the MC implements the evolution of the number density $D(x)/x$. This algorithm can be cross-checked against the time evaluation in \cref{eq:cascadePS_exp}, and the numerical evolution of \cref{eq:cascade}.

\begin{figure}
    \centering
    \includegraphics[width=\linewidth]{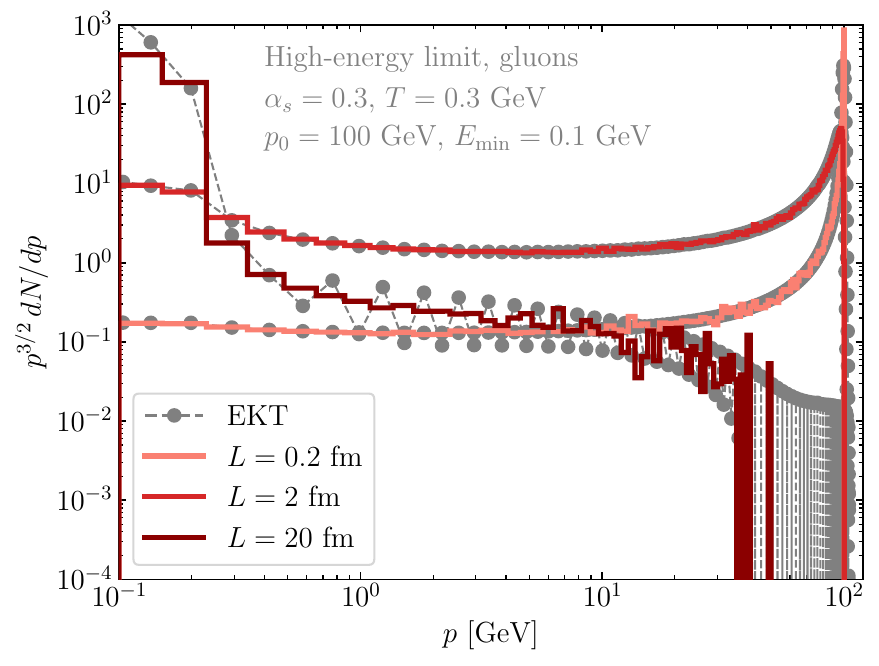}
    \caption{The energy distribution in the high-energy limit, evaluated for different path lengths with two different implementations.}
    \label{fig:SoftShower}
\end{figure}

We evaluated \cref{eq:cascade} using a differential equation solver and \cref{eq:cascadePS} with a parton shower. We use the high-energy splitting rate ($T\ll p$), $\Gamma(z,\p)=\frac{\alpha_s}{2\pi}\frac{C_A}{z(1-z)}\sqrt{\frac{C_A\hat q}{z(1-z)p}}$, where $\hat q=\frac{g^2C_ATm_D^2}{2\pi}$, and a single 100 GeV gluon as the initial condition. This splitting rate is the usual choice of medium cascades, and it reproduces the EKT rates in the deep LPM limit (up to logarithmic corrections to $\hat q$). This splitting rate has an infrared pole, and thus the separate terms in \cref{eq:cascade,eq:cascadePS} are divergent. The overall sum is, however, finite as divergences cancel in gain and loss terms. Therefore, we introduced an infrared regulator for both algorithms such that every splitting involves particles with energies bigger than $E_{\min}$. For not too late times, distributions far from $E_{\min}$ are independent of the cutoff.\footnote{Common Boltzmann equation solvers implicitly regulate infrared divergences through discrete and finite momentum grids.} The number density is evaluated through the histogram
\begin{equation}\label{eq:Histogram_1d}
    \frac{\rmd N}{\rmd p} = \frac{1}{N_{\rm ev}}\sum_{n=1}^{N_{\rm ev}}\sum_{i=1}^{N_n}\frac{w_i}{\rmd p}\,,
\end{equation}
where $N_{\rm ev}$ is the number of events, $N_n$ is the number of final partons in the event, and $w_i$ is the weight of the parton ($w_i=1$ for now).

\Cref{fig:SoftShower} shows our results with the two different algorithms for different propagation lengths, gray for the traditional diffeq. solver, and color for the new MC. Numerical uncertainties are present for the longest propagation (oscillating gray points). For short paths, most of the energy is in the initial perturbation, while later the cascade quickly transfers energy to the lowest modes. At late time the initial peak disappears and soft modes pile up at the infrared cutoff. Varying $E_{\min}$ changes the late-time position of the peak. Since in the high-energy limit we neglected merging processes, detailed balance is broken and the gluon distribution does not thermalize. In the next section, we derive the full EKT evolution, which includes both splittings and mergings, recovering thermalization at late times.

\section{Thermalization of medium induced shower}
Returning to the full EKT evolution, \cref{eq:PartonShower}, one proceeds analogously to the high-energy limit, but now accounting for multiple competing channels of splitting, merging, and scattering. In this work, we consider gluons and, for simplicity, neglect elastic collisions. Under this assumption, the inelastic dynamics effectively reduce to two processes in \cref{eq:linC12}:
\begin{equation}
    \begin{split}
        D(t,x;p_0) & = \Delta^{1\leftrightarrow2}(t,xp_0) D(t_0,x;p_0)                         \\
                   & + \int_{t_0}^t\rmd t'\Delta^{1\leftrightarrow2}(t-t', xp_0)\int_0^1\rmd z \\
                   & \times\Big[2\Gamma_1(z,\tfrac{xp_0}{z}) D(t',\tfrac{x}{z};p_0)            \\
                   & \quad- \Gamma_2(z,\tfrac{\bar z xp_0}{z}) D(t',\tfrac{\bar{z}x}{z};p_0)   \\
                   & \quad+ \Gamma_2(z,\bar z xp_0) D(t',\bar{z}x;p_0)\Big]\,.
    \end{split}
\end{equation}
The Sudakov factor is
\begin{equation}
    \ln\Delta^{1\leftrightarrow2}(t,p)=-\int_{t_0}^t \rmd t'\,\rmd z\,(\Gamma_1(z,p)+\Gamma_2(z,p))\,,
\end{equation}
with the corresponding real emission processes:
\begin{enumerate}
    \item[(1)] $g(p)\to g(zp)+g(\bar zp)$, with rate
          \begin{equation}
              \Gamma_1(z,p)=\tfrac{1}{2}\Gamma(z,p)(1+n(zp)+n(\bar zp))\,,
          \end{equation}
          where $\bar z\!=\!1-z$, and splitting always dominates over merging.
    \item[(2)] $g(p)+g(zp/\bar z)\leftrightarrow g(p/\bar z)$, with rate
          \begin{equation}
              \Gamma_2(z,p)=\tfrac{1}{\bar z^3}\Gamma(z,\tfrac{p}{\bar z})(n(\tfrac{zp}{\bar z})-n(\tfrac{p}{\bar z}))\,,
          \end{equation}
          which allows both gain and loss contributions (positive and negative terms). The loss contribution is accounted for by introducing a label that distinguishes whether a particle is a hole or not. Hole particles split into hole daughters, while the merging of two holes produces a normal particle. The mixing is also possible by following $g+g\leftrightarrow g$ with their respective signs. When particles are fed to e.g. hadronization, labels are inherited to the offspring. When final particles are analyzed, histogram entries can receive positive or negative contributions depending on whether the particle is a hole.
\end{enumerate}
The splitting kernels have infrared poles, which are regulated by introducing an energy cutoff, requiring the energy of all splitting legs to be larger than $E_{\rm min}$. Other regulator definitions are also possible e.g. regulating $z$. The resulting $\delta f(p)$ should be reliable far from the cutoff $p\gg E_{\min}$. We postpone further studies of infrared sensitivity in EKT to upcoming works.

In the parton shower, the next splitting time is sampled from the Sudakov factor $\Delta^{1\leftrightarrow2}(t_{\rm next},p_0)$. Once a branching occurs ($t_{\rm next}<L$), one of the two channels is chosen in proportion to their total rate, and the momentum fraction $z$ is drawn accordingly from $\Gamma_i(z,p_0)$. This is analogous to competing flavor channels in jet showers~\cite{Lonnblad:2012hz}. The procedure is iterated for all new particles until $t_{\rm next}>L$. Finally, a Sudakov factor is sampled to account for the absence of interactions in the interval $t_{\rm last}<L$ to $L$. When reconstructing the momentum, one has to be careful if the splitting involves holes. The number density is evaluated as in \cref{eq:Histogram_1d}.

\begin{figure}
    \centering
    \includegraphics[width=\linewidth]{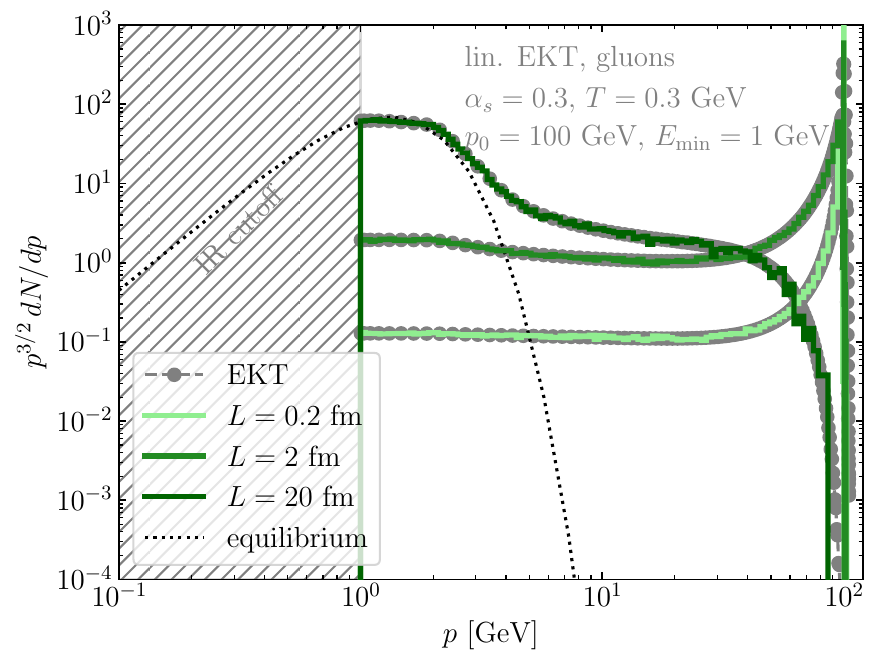}
    \caption{Energy distribution for different path lengths, comparing the EKT solver and the parton-shower implementation.}
    \label{fig:FullShower}
\end{figure}

\Cref{fig:FullShower} compares the numerical evaluation of \cref{eq:PartonShower} using the EKT differential-equation solver with the new MC parton shower. Both approaches yield identical distributions. For clarity, we employed the same high-energy splitting rates and neglected elastic collisions, leaving their numerical implementation for future work. 
We recover a thermal distribution near $p\!\sim\!T$ characterizing the equilibration of the perturbation~\cite{Schlichting:2020lef}. Excessive splittings and mergings at low momenta lead to a large number of low-energy particles, slowing down the simulation. The relatively large infrared cutoff $E_{\min}\!=\!1$ GeV avoids this issue, while still allowing for a clear thermal peak.\footnote{
    Generally speaking, the peak arises from a combination of the thermal bump and the pile-up of particles with $p\sim E_{\min}$. We numerically verified that for $E_{\min}\lesssim T$ only the thermal peak contributes.
}
This value is lower than the usual cutoff, $\approx\!5$ GeV, used in other works. Such cutoffs are common in phenomenological applications, as below $\sim\!1$~GeV, non-perturbative and hadronization effects become important. Differential equation solvers on the other hand use much lower cutoffs i.e. $E_{\min}\ll T$. For reference, the asymptotic late-time distribution, $\delta f_{\rm eq}(\p,\x)=\delta T\partial_T n_B(p)$, is also shown with a dotted line, where $\frac{p_0}{V}=\frac{4\pi^2}{30}T^3\delta T$.
There are clear deviations from equilibrium at intermediate times illustrating the need for out-of-equilibrium kinetic evolution instead of instantaneous thermalization and hydro wakes in phenomenological applications.

\begin{figure}
    \centering
    \includegraphics[width=\linewidth]{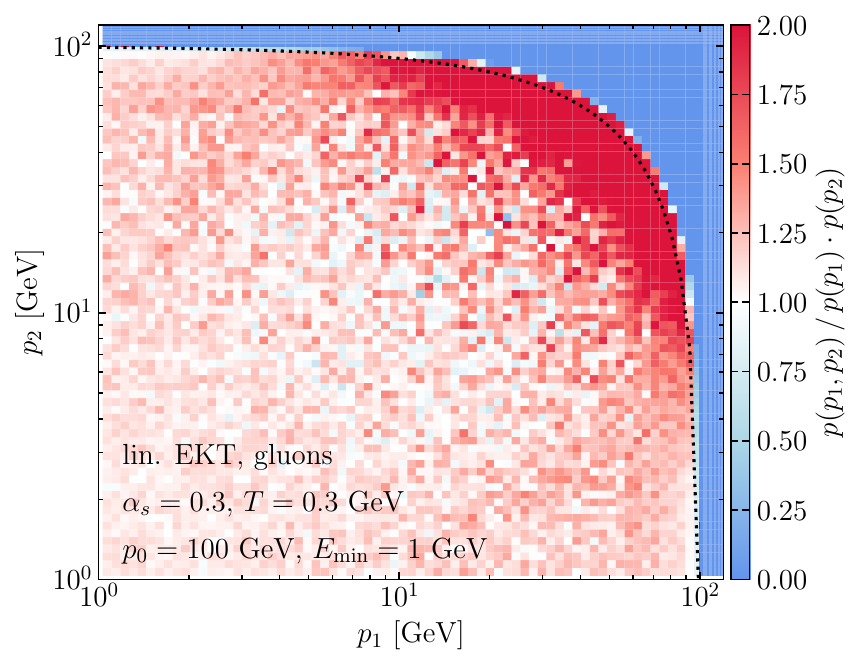}
    \caption{Two-particle energy distribution relative to molecular chaos after $L=2$ fm propagation.}
    \label{fig:FullShower_corr}
\end{figure}

Finally, a unique application of the new algorithm is evaluating $n$-particle distributions and thus correlations. For example, we compute the two-particle distribution as
\begin{equation}\label{eq:Histogram_2d}
    \frac{\rmd N}{\rmd p_1\rmd p_2}=\frac{1}{N_{\rm ev}}\sum_{n=1}^{N_{\rm ev}}\sum_{i\neq j=1}^{N_n}\frac{w_iw_j}{\rmd p_1\rmd p_2}\,.
\end{equation}
We show \cref{eq:Histogram_2d} in \cref{fig:FullShower_corr}, divided by two single-particle distributions representing trivial correlations. We normalize all distributions to 1 before taking the ratio, as events typically produce many particles. \Cref{eq:Histogram_2d} shows a clear boundary originating from energy conservation $\sum_ip^0_i=p_0$, in the figure, $p_1+p_2=p_0$ is shown with a dotted line.
Furthermore, additional correlations are present from two sources: multi-parton final states, and their origin from inelastic collisions. In the limit $p_1\ll p_2\approx p_0$ (and $p_2\ll p_1\approx p_0$), the ratio approaches to 1. With \cref{fig:FullShower_corr}, we illustrate that our parton shower clearly deviates beyond the usual assumption of molecular chaos of the Boltzmann equation, where the two-particle distribution are the product of two single particle distributions, i.e.~$\delta f(p_1,p_2)\approx\delta f(p_1)\cdot\delta f(p_2)$.

We note that the parton shower algorithm not only provides multi particle states but also predicts fluctuations through its stochastic nature, while conserving energy and momentum. Introducing fluctuations into effective theories such as stochastic hydrodynamics and understanding thermal-like fluctuations out of equilibrium are of great interest in heavy-ion collisions, with a potentially significant impact on small collision systems~\cite{Kapusta:2011gt,Basar:2024srd,SoaresRocha:2024afv}.

\section{Conclusions}

We have reformulated the linearized QCD effective kinetic theory (EKT) as a parton shower and benchmarked it against established numerical results. This framework provides, on the one hand, a general prescription for implementing any linearized kinetic theory as a ``quasi-particle shower'', and on the other, a method for thermalizing jet-quenching Monte Carlo parton showers by incorporating statistical factors that ensure detailed balance. Within this formulation, splittings, mergings, recoils, and hole excitations acquire a transparent interpretation in terms of gain and loss processes. By studying the energy distribution, we demonstrated both the turbulent cascade and the approach to thermal equilibrium in the linearized EKT and in the corresponding shower algorithm. In this first work, we focused on gluons and inelastic collisions, leaving the inclusion of quarks and elastic processes for future studies.

Looking ahead, our algorithm evolves partons on an event-by-event basis, enabling straightforward interfaces with vacuum jet showers, hadronization models, and other tools relevant to jet-quenching phenomenology. Beyond reproducing single-particle distributions, parton showers naturally provide access to multi-particle correlations, offering new opportunities for EKT-based investigations. We presented early results for two-particle correlations that go beyond the usual assumption of molecular chaos. Furthermore, this framework can be extended to study fluctuations and space-time inhomogeneities, which are currently neglected in EKT. We therefore anticipate that this approach will significantly broaden the reach of current simulations, with potential applications ranging from jet quenching to the general study of out-of-equilibrium dynamics.

\begin{acknowledgements}
    We greatly appreciate discussions with Aleksas Mazeliauskas and Daniel Pablos. The work of A.T. is supported by DFG through Emmy Noether Programme (project number 496831614) and through CRC 1225 ISOQUANT (project number 27381115).
    IS is indebted to K.~Eskola, H.~Niemi for fruitful discussions throughout the evolution of this project.
    IS was funded as a part of the European Research Council project ERC-2018-ADG-835105 YoctoLHC, and as a part of the Center of Excellence in Quark Matter of the Academy of Finland (Projects No. 346325 and 364192).
\end{acknowledgements}

\bibliography{ref_vac,ref_med}

\end{document}